\documentclass[prl,twocolumn,showpacs,preprintnumbers,amsmath,amssymb,superscriptaddress]{revtex4}

\usepackage{graphicx}
\usepackage{dcolumn}
\usepackage{bm}

\begin{document}

\title{Collective quantum jumps of Rydberg atoms}

\author{Tony E. Lee}
\affiliation{Department of Physics, California Institute of Technology, Pasadena, California 91125, USA}
\author{H. H\"{a}ffner}
\affiliation{Department of Physics, University of California, Berkeley, California 94720, USA}
\author{M. C. Cross}
\affiliation{Department of Physics, California Institute of Technology, Pasadena, California 91125, USA}

\date{\today}

\begin{abstract}
We study an open quantum system of atoms with long-range Rydberg interaction, laser driving, and spontaneous emission. Over time, the system occasionally jumps between a state of low Rydberg population and a state of high Rydberg population. The jumps are inherently collective and in fact exist only for a large number of atoms. We explain how entanglement and quantum measurement enable the jumps, which are otherwise classically forbidden.
\end{abstract}

\maketitle
Perhaps the strangest aspect of quantum mechanics is the notion that merely observing a system changes it. This concept is taken to the extreme in the quantum Zeno effect, where the constant observation of a system inhibits a transition that would otherwise take place \cite{itano90}. Another equally striking phenomenon is quantum jumps, where a system under continuous monitoring occasionally switches between two distinct states \cite{cook85,plenio98}. Quantum jumps have been observed in many settings, such as trapped ions \cite{nagourney86,sauter86,bergquist86}, photons \cite{gleyzes07}, electrons \cite{vamivakas10}, and superconducting qubits \cite{vijay11}. In these experiments, the object being observed is a single particle or can be described by a single degree of freedom. But the recent interest in generating multi-particle entanglement \cite{leibfried05,haeffner05,monz11} raises the question of how \emph{large} systems of entangled particles behave under constant observation. For example, do they undergo collective quantum jumps?

In this paper, we show how entanglement and quantum measurement lead to collective quantum jumps of Rydberg atoms. A Rydberg atom is an atom excited to a high energy level $n$. The dipole-dipole interaction between two Rydberg atoms is strong and allows one to entangle many atoms over long distances \cite{lukin01}. This interaction has attracted recent interest for quantum-information processing \cite{jaksch00,lukin01,wilk10,isenhower10,weimer10,saffman10} and many-body physics \cite{pohl10,lesanovsky10,cinti10,honer10,ji11,lee11b}.

We consider a group of atoms laser-driven to the Rydberg state and spontaneously decaying back to the ground state. Classical mean-field theory predicts two stable collective states, one with low Rydberg population and one with high Rydberg population. Classically, the system should remain in one of the stable states. However, we find that quantum fluctations drive transitions between the states, resulting in quantum jumps. The jumps are inherently collective and exist only for a large number of atoms. Our results may be extended to other settings, such as coupled optical cavities \cite{gerace09,carusotto09,hartmann10,tomadin10} and quantum-reservoir engineering \cite{diehl08,diehl10,tomadin11,verstraete09}.

Two atoms in the same Rydberg level experience an energy shift $V$ due to their dipole-dipole interaction \cite{saffman10}. The dependence of $V$ on inter-particle distance $R$ can take several forms. In the presence of a static electric field, $V\sim 1/R^3$ and is anisotropic. In the absence of a static field, $V\sim 1/R^3$ for small distances and $1/R^6$ for large distances, and the interaction can be isotropic or anisotropic, depending on the Rydberg level. In this paper, we are interested in the long-range type of coupling ($V\sim 1/R^3$). However, to be able to simulate large systems, we approximate the long-range coupling as a constant all-to-all coupling with suitable normalization; this approximation is appropriate for a two or three-dimensional lattice for the system sizes used here.

Consider a system of $N$ atoms continuously excited by a laser from the ground state to a Rydberg state. Let $|g\rangle_j$ and $|e\rangle_j$ denote the ground and Rydberg states of atom $j$. The Hamiltonian in the interaction picture and rotating-wave approximation is ($\hbar=1$)
\begin{eqnarray}
H&=&\sum_j \left[-\Delta \,|e\rangle\langle e|_j+\frac{\Omega}{2}(|e\rangle\langle g|_j+|g\rangle\langle e|_j) \right]\nonumber\\
&& + \frac{V}{N-1}\sum_{j<k}\,|e\rangle\langle e|_j\otimes |e\rangle\langle e|_k \;,\label{eq:H}
\end{eqnarray}
where $\Delta=\omega_\ell-\omega_o$ is the detuning between the laser and transition frequencies and $\Omega$ is the Rabi frequency, which depends on the laser intensity.

The Rydberg state has a finite lifetime due to spontaneous emission and blackbody radiation. When an atom spontaneously decays from the Rydberg state, it usually goes directly to the ground state or first to a low-lying state \cite{gallagher94}; since the low-lying states have relatively short lifetimes, we ignore them. In addition, blackbody radiation may transfer an atom from a Rydberg level to nearby levels, but this is minimized by working at cryogenic temperatures \cite{beterov09}. Thus, each atom is approximated as a two-level system, and we account for spontaneous emission from the Rydberg state using the linewidth $\gamma$ \cite{lee11b}. Note that each atom emits into different electromagnetic modes due to the large inter-particle distance.

The environment absorbs all the spontaneously emitted photons, so the atoms are continuously monitored by the environment. We are interested in the temporal properties of the emitted photons. There are two equivalent ways to study such an open quantum system. The first is the master equation, which describes how the density matrix of the atoms, $\rho$, evolves in time:
\begin{eqnarray}
\dot{\rho}&=&-i[H,\rho]\nonumber\\
&&+\gamma\sum_j\left(-\frac{1}{2}\{|e\rangle\langle e|_j,\rho\}+|g\rangle\langle e|_j\,\rho\,|e\rangle\langle g|_j\right)\;.\label{eq:master}
\end{eqnarray}
A master equation of this form has a unique steady-state solution \cite{schirmer10}, $\rho_{ss}$, which can be found numerically by Runge-Kutta integration. The integration can be vastly sped up by utilizing the fact that the atoms are symmetric under interchange due to all-to-all coupling; the complexity is then $O(N^3)$ instead of $O(4^N)$. Using $\rho_{ss}$, one can calculate the statistics of the emitted light. In particular, the correlation of photons emitted by two different atoms is $g^{(2)}_{ij}=\langle E_i E_j \rangle/\langle E_i \rangle\langle E_j \rangle$,
where $E_i\equiv|e\rangle\langle e|_i$ \cite{scully97}. If $g^{(2)}_{ij}>1$, the atoms tend to emit in unison (bunching); if $g^{(2)}_{ij}<1$, they avoid emitting in unison (antibunching).

The second approach is the method of quantum trajectories, which simulates how the wave function evolves in a single experiment \cite{dalibard92,molmer93,dum92}. In the simulation, the environment observes at every time step whether an atom has emitted a photon, and the wavefunction is updated accordingly; the crucial point is that even when no photon is detected, the wave function is still modified. The algorithm is as follows. Given the wave function $|\psi(t)\rangle$, one randomly decides whether an atom emits a photon in the time interval $[t,t+\delta t]$ based on its current Rydberg population. If atom $j$ emits a photon, the wave function is collapsed: $|\psi(t+\delta t)\rangle=|g\rangle\langle e|_j |\psi(t)\rangle$. If no atoms emit a photon, $|\psi(t+\delta t)\rangle=(1-iH_{\mbox{\scriptsize{eff}}}\delta t)|\psi(t)\rangle$, where $H_{\mbox{\scriptsize{eff}}}=H-(i\gamma/2)\sum_j |e\rangle\langle e|_j$. After normalizing the wave function, the process is repeated for the next time step. The nonunitary part of $H_{\mbox{\scriptsize{eff}}}$ is a shortcut to account for the fact that \emph{the non-detection of a photon shifts the atoms toward the ground state} \cite{dalibard92,molmer93}.

These two approaches are related: the master equation describes an ensemble of many individual trajectories \cite{dalibard92,molmer93}. Also, $\rho_{ss}$ can be viewed as the ensemble of wave functions that a single trajectory explores over time. We will use both approaches below, although quantum jumps are most clearly seen using quantum trajectories.

\begin{figure}
\centering
\includegraphics[width=3.5 in]{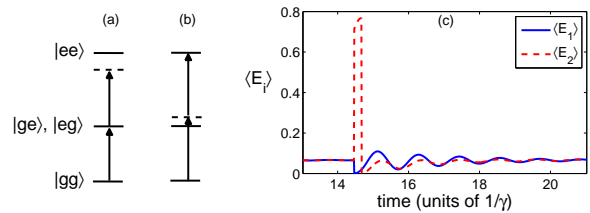}
\caption{\label{fig:n2_qtraj}Two atoms. (a) When $\Delta=0$, $|ee\rangle$ is uncoupled from the other states. (b) When $\Delta=V/2$, there is a resonant two-photon transition between $|gg\rangle$ and $|ee\rangle$. (c) Quantum trajectory simulation with $\Omega=1.5\gamma$, $\Delta=V/2=5\gamma$, showing Rydberg population of each atom over time. Atom 1 (solid blue line) emits at $t=14.4/\gamma$, which causes $\langle E_2\rangle$ (dashed red line) to suddenly increase. Atom 2 then emits at $t=14.7/\gamma$. When no photons have been emitted for a while, the wave function approaches a steady state.}
\end{figure}

\begin{figure}
\centering
\includegraphics[width=3.5 in]{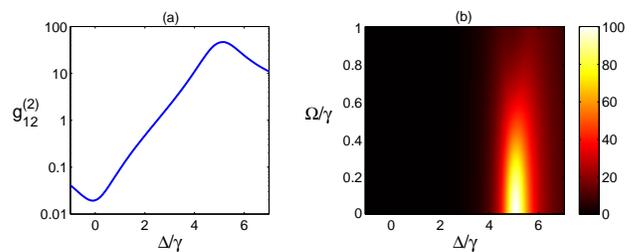}
\caption{\label{fig:n2_g2}Photon correlation for two atoms with $V=10\gamma$. (a) Correlation vs $\Delta$ for $\Omega=0.5\gamma$. (b) Correlation as a function of $\Omega$ and $\Delta$ using color scheme on right.}
\end{figure}

We first consider the case of $N=2$ atoms since it is instructive for larger $N$. Laser excitation and spontaneous emission distribute population through out the Hilbert space, $\{|gg\rangle,|ge\rangle,|eg\rangle,|ee\rangle\}$. When $\Delta=0$, $|ee\rangle$ is uncoupled from the rest of the states due to its energy shift, so there is little population in it [Fig.~\ref{fig:n2_qtraj}(a)]; this is the well-known blockade effect \cite{lukin01,saffman10}. But when $\Delta\approx V/2$, there is a resonant two-photon transition between $|gg\rangle$ and $|ee\rangle$, so $|ee\rangle$ becomes populated [Fig.~\ref{fig:n2_qtraj}(b)]. Using the master equation, one can calculate the photon correlation between the two atoms (Figs.~\ref{fig:n2_g2}). There is strong antibunching for $\Delta\approx 0$ and strong bunching for $\Delta\approx V/2$, which makes sense since a joint emission requires population in $|ee\rangle$.
One can solve for the correlation as a perturbation series in $\Omega$:
\begin{eqnarray}
g^{(2)}_{12}=\frac{1+4\Delta^2}{1+(V-2\Delta)^2} + \frac{4V(V-4\Delta)}{[1+(V-2\Delta)^2]^2}\Omega^2 + O(\Omega^4) \;.
\end{eqnarray}
Note that the correlation can be made arbitrarily large by setting $\Omega\approx 0$, $\Delta=V/2$, and $V$ large; this may be useful as a heralded single-photon source \cite{eisaman11}.

Further insight is provided by quantum trajectories. An example trajectory for $\Delta=V/2$ is shown in Fig.~\ref{fig:n2_qtraj}(c). The atoms emit photons at various times. When no photons have been emitted for a while, the wave function approaches an entangled steady state due to the balance of laser excitation and nonunitary decay from the non-detection of photons \cite{note1}:
\begin{eqnarray}
|\psi\rangle_{ss}&=&c_1|gg\rangle+c_2|ge\rangle+c_3|eg\rangle+c_4|ee\rangle\;,\label{eq:psi_n2_before}
\end{eqnarray}
where the coefficients have constant magnitudes and their phases evolve with the same frequency (this is a periodic steady state). Because of the laser detuning, $|c_1|^2$ is much larger than $|c_2|^2,|c_3|^2,|c_4|^2$, which are comparable to each other. Thus $\langle E_1\rangle,\langle E_2\rangle\approx0$ and the atoms are unlikely to emit. But when atom 1 happens to emit, the wave function becomes
\begin{eqnarray}
|\psi\rangle=\frac{c_3|gg\rangle+c_4|ge\rangle}{|c_3|^2+|c_4|^2}\;,\label{eq:psi_n2_after}
\end{eqnarray}
Now, $\langle E_2\rangle$ is large and atom 2 is likely to emit, which leads to photon bunching [Fig.~\ref{fig:n2_qtraj}(c)]. 

\begin{figure}
\centering
\includegraphics[width=3.5 in]{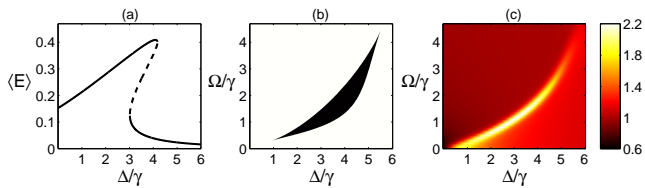}
\caption{\label{fig:bistability}(a) Fixed points of mean-field model as function of detuning for $\Omega=1.5\gamma$ and $V=10\gamma$. Stable (unstable) fixed points are denoted by solid (dashed) lines. (b) Mean-field bistable region (black) for $V=10\gamma$. (c) Photon correlation $g^{(2)}_{ij}$ for 16 atoms with same parameters as (b), using color scheme on right.}
\end{figure}

Then we consider the case of large $N$. We first review mean-field theory, since it is important for what follows \cite{hopf84,lee11b}. Mean-field theory is a classical approximation to the quantum model: correlations between atoms are ignored, and the density matrix factorizes by atom, $\rho=\bigotimes_{j=1}^N\overline{\rho}$, where $\overline\rho$ evolves according to
\begin{eqnarray}
\dot{\overline\rho}_{ee}&=&-\Omega\,\mbox{Im }\overline\rho_{eg}-\gamma\overline\rho_{ee} \label{eq:w}\;,\\
\dot{\overline\rho}_{eg}&=&i(\Delta-V\overline\rho_{ee})\overline\rho_{eg}-\frac{\gamma}{2}\overline\rho_{eg}+i\Omega\left(\overline\rho_{ee}-\frac{1}{2}\right) \label{eq:q}\;.
\end{eqnarray}
These are the optical Bloch equations for a two-level atom, except the effective laser detuning is $\Delta-V\overline\rho_{ee}$. There are one or two stable fixed points, depending on the parameters (Fig.~\ref{fig:bistability}). Classically, the system should go to a stable fixed point and stay there, since there are no other attracting solutions.

\begin{figure}
\centering
\includegraphics[width=3.5 in]{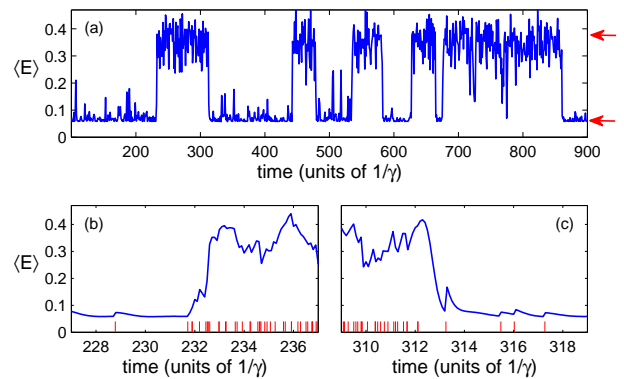}
\caption{\label{fig:n16_qtraj}Quantum trajectory of 16 atoms showing average Rydberg population over time with $\Omega=1.5\gamma$, $V=10\gamma$, $\Delta=3.4\gamma$. (a) Quantum jumps between two metastable collective states. Red arrows point at the stable fixed points of mean-field theory. (b) and (c) are zoomed-in views, and red lines mark photon emissions. (b) Rapid succession of emissions around $t=232/\gamma$ causes a jump up. (c) Absence of emissions around $t=313/\gamma$ causes a jump down.}
\end{figure}

Now we consider the original quantum model for large $N$. Figure \ref{fig:n16_qtraj}(a) shows a quantum trajectory for $N=16$. We plot the average Rydberg population of all the atoms $\langle E\rangle$, where $E\equiv\sum_i E_i/N$. $\langle E\rangle$ appears to switch in time between two values. In fact, these two values correspond to the two stable fixed points of mean-field theory for the chosen parameters. Thus, we find that the quantum model is able to jump between the stable states of the classical mean-field model. When the parameters are such that mean-field theory is monostable, $\langle E\rangle$ remains around one value and there are no jumps. As a result, the photons are bunched when mean-field theory is bistable and uncorrelated otherwise. This correspondence is evident in Fig.~\ref{fig:bistability}(b)-(c), with better agreement for larger $N$.



We call the two states in Fig.~\ref{fig:n16_qtraj}(a) the dark and bright states, since the one with lower $\langle E\rangle$ has a lower emission rate. In the dark state, the wave function approaches a steady state, $|\psi\rangle_{ss}$, in between the sporadic emissions. This is due to the balance of laser excitation and non-unitary decay from the non-detection of photons, similar to the case of two atoms. In the bright state, the large Rydberg population brings the system effectively on resonance. The bright state sustains itself because an atom is quickly reexcited after emitting a photon.

\begin{figure}
\centering
\includegraphics[width=3.5 in]{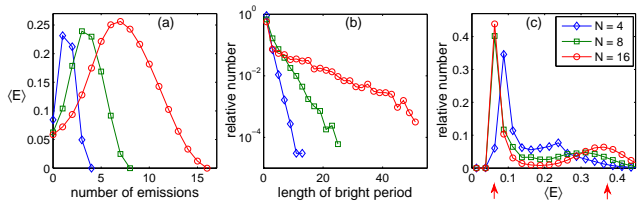}
\caption{\label{fig:stats}Statistics for $\Omega=1.5\gamma$, $\Delta=3.5\gamma$, $V=10\gamma$ comparing $N=4,8,16$. (a) Rydberg population $\langle E\rangle$ after a number of simultaneous emissions from the steady-state wave function $|\psi\rangle_{ss}$. (b) Length distribution of bright periods (in units of $1/\gamma$), using arbitrary threshold of $\langle E\rangle=0.2$ and sampling rate of $10\gamma$. (c) Distribution of $\langle E\rangle$. Red arrows point at stable fixed points of mean-field theory.}
\end{figure}

Suppose the system is in the dark state. The steady-state wavefunction $|\psi\rangle_{ss}$ is an entangled state of all the atoms with most population in $|gg\ldots g\rangle$. Although $\langle E\rangle$ is small, when an atom happens to emit a photon, $\langle E\rangle$ increases due to the entangled form of $|\psi\rangle_{ss}$. In fact, if more atoms emit within a short amount of time, $\langle E\rangle$ increases further [Fig.~\ref{fig:stats}(a)]. When enough atoms have emitted such that $\langle E\rangle$ is high, the system is in the bright state and sustains itself there [Fig.~\ref{fig:n16_qtraj}(b)]. If too few atoms emitted, the system quickly returns to $|\psi\rangle_{ss}$.

Then suppose the system is in the bright state. There are two ways to jump to the dark state: most of the atoms emit simultaneously or most of the atoms do not emit for a while (the non-detection of photons projects the atoms toward the ground state). For our parameters, simulations indicate that the latter is usually responsible for the jumps down [Fig.~\ref{fig:n16_qtraj}(c)].

The jumps are inherently collective, since they result from joint emissions or joint non-emissions. As $N$ increases, the dark and bright periods become longer and more distinct [Fig.~\ref{fig:stats}(b)-(c)]. This can be understood intuitively as follows. Suppose the system is in $|\psi\rangle_{ss}$. As $N$ increases, the increment of $\langle E\rangle$ per emission decreases [Fig.~\ref{fig:stats}(a)]. Thus, for large $N$, a rapid succession of many emissions is necessary to jump to the bright state. Although the emission rate in the dark state increases with $N$, the rate of nonunitary decay in $H_{\mbox{\scriptsize{eff}}}$ also increases with $N$. The result is that the probability rate of a jump up decreases. Then suppose the system is in the bright state. As $N$ increases, a jump down requires more atoms to not emit in some time interval, so the probability rate of a jump down decreases.


These collective jumps are reminiscent of a familiar classical effect. It is well known that adding thermal noise to a bistable classical system induces transitions between the two stable fixed points \cite{dykman79,aldridge05}. In contrast, the jumps here are induced by \emph{quantum} noise due to entanglement and quantum measurement. We note that the jumps may be the many-body version of quantum activation, in which quantum fluctuations drive transitions over a classical barrier \cite{marthaler06,katz07}.




Experimentally, the jumps may be observed in a 2D optical lattice of atoms with a static electric field normal to the plane for long-range Rydberg interaction. For example, two ${}^{87}\mbox{Rb}$ atoms in the $|n=15,q=14,m=0\rangle$ Rydberg state have a coupling of about $44 \mbox{ kHz}$ at a distance of $13 \mbox{ }\mu\mbox{m}$ \cite{jaksch00}, and the linewidth is $\gamma/2\pi\approx 68\mbox{ kHz}$ at 0 K \cite{beterov09}. This corresponds to $V\approx10\gamma$ with $N=16$ atoms for the all-to-all model in Eq.~\eqref{eq:H}, which are the parameters used in our discussion. One could observe the jumps directly by monitoring the fluorescence from the atoms. Alternatively, one could make repeated projective measurements and thereby infer the existence of two metastable states from the distribution of $\langle E\rangle$.


Thus, atoms coupled through the Rydberg interaction exhibit collective quantum jumps. It would be interesting to see whether similar jumps appear in other settings, such as coupled optical cavities \cite{gerace09,carusotto09,hartmann10,tomadin10} and quantum-reservoir engineering \cite{diehl08,diehl10,tomadin11,verstraete09}. In particular, since mean-field bistability seems to predict collective jumps in the underlying quantum model, one should look for bistability in the mean-field models of other systems \cite{diehl10,tomadin11,tomadin10}.  Finally, we note that one can observe the conventional type of quantum jump in a three-level atom by using the Rydberg level as the metastable state \cite{cook85}.  Then due to the Rydberg interaction, a jump in one atom will enhance or inhibit jumps in its neighbors. This may lead to interesting spatiotemporal dynamics.

We thank G. Refael and R. Lifshitz for useful discussions. This work was supported by NSF Grant No. DMR-1003337.

\bibliography{jump}

\end{document}